\PassOptionsToPackage{table}{xcolor}
\documentclass[acmsmall]{acmart}

\usepackage{needspace}
\usepackage{url}
\usepackage{adjustbox}
\usepackage{algorithm}
\usepackage{algpseudocode}
\usepackage{multirow}
\usepackage{tikz}
\usepackage{amsmath}
\usepackage{xspace}
\usepackage{enumitem}
\usepackage{booktabs}
\usepackage{bbding}
\usepackage[caption=false]{subfig}
\usepackage{stfloats}
\usepackage{indentfirst}
\usepackage{makecell}
\usepackage{tcolorbox}
\usepackage{lipsum}
\usepackage{minted}
\usepackage{graphicx}
\usepackage[normalem]{ulem}
\usepackage{multicol}
\usepackage{array}
\usepackage{ragged2e}
\usepackage{tabularx}
\usepackage{soul}

\newcommand{\ie}{{\em i.e.},\xspace}
\newcommand{\eg}{{\em e.g.},\xspace}

\newcommand{\vulkey}{\textsc{Vul}\textsc{Key}\xspace}

\tcbuselibrary{listingsutf8}
\tcbuselibrary{skins}
\tcbset{
  mymintedbox/.style={
    colback=gray!20,
    colframe=gray!50,
    boxrule=0.5pt,
    arc=3pt,
    boxsep=0pt,
    left=2pt,
    right=2pt,
    top=1pt,
    bottom=1pt,
    fontupper=\ttfamily\footnotesize,
    listing options={breaklines=true, breakautoindent=true, breakindent=2em}
  }
}

\AtBeginDocument{%
  }


\setcopyright{cc}
\setcctype{by}
\acmDOI{10.1145/3808117}
\acmYear{2026}
\acmJournal{PACMSE}
\acmVolume{3}
\acmNumber{FSE}
\acmArticle{FSE110}
\acmMonth{7}
\acmSubmissionID{fse26mainb-p225-p}
\received{2025-09-12}
\received[accepted]{2026-03-24}

\begin{document}

\title{VulKey: Automated Vulnerability Repair Guided by Domain-Specific Repair Patterns}

\author{Jia Li}
\orcid{0009-0007-5557-4916}
\affiliation{%
  \institution{The Chinese University of Hong Kong}
  \city{Hong Kong}
  \country{China}
}
\email{linsayli@link.cuhk.edu.hk}

\author{Zhuangbin Chen}
\authornote{Corresponding author.}
\orcid{0000-0001-5158-6716}
\affiliation{%
  \institution{Sun Yat-sen University}
  \city{Zhu Hai}
  \country{China}
}
\email{chenzhb36@mail.sysu.edu.cn}

\author{Yuxin Su}
\orcid{0000-0002-3338-8561}
\affiliation{%
  \institution{Sun Yat-sen University}
  \city{Zhu Hai}
  \country{China}
}
\email{suyx35@mail.sysu.edu.cn}

\author{Michael R. Lyu}
\orcid{0000-0002-3666-5798}
\affiliation{%
  \institution{The Chinese University of Hong Kong}
  \city{Hong Kong}
  \country{China}
}
\email{lyu@cse.cuhk.edu.hk}

\begin{abstract}
The increasing prevalence of software vulnerabilities highlights the need for effective Automatic Vulnerability Repair (AVR) tools. While LLM-based approaches are promising, they struggle to incorporate structured security knowledge from sources like CWE and NVD. Current methods either use this information superficially by concatenating the CWE-ID into the input prompt, yielding negligible benefits, or rely on few-shot learning with rigid, non-generalizable examples, which limits their effectiveness in real-world scenarios.
To address this gap, we propose \vulkey, an LLM-based AVR framework that leverages a hierarchical abstraction of expert knowledge to guide patch generation. Our novel three-level abstraction formulates repair strategies in terms of \textit{CWE type}, \textit{syntactic actions}, and \textit{semantic key elements}. This approach captures the essence of a security fix with greater generality than concrete examples and more semantic richness than traditional syntax-based templates, overcoming the coverage limitations of prior methods.
\vulkey is implemented as a two-stage pipeline: first, expert knowledge matching predicts an appropriate repair pattern for the vulnerability; second, repair code generation uses a pattern-guided, fine-tuned LLM to produce secure patches. 
On the real-world C/C++ dataset \textit{PrimeVul}, \vulkey achieves 31.5\% repair accuracy, surpassing the best baseline by 7.6\% and outperforming leading tools such as \textsc{VulMaster} and \textsc{GPT-5}. Moreover, \vulkey demonstrates cross-language and cross-model generalizability, with state-of-the-art performance on the Java benchmark \textit{Vul4J}. These results underscore the importance of structured expert knowledge in advancing AVR effectiveness.
Our work demonstrates that explicitly modeling and integrating expert security knowledge through hierarchical patterns is a crucial step toward building more effective and reliable AVR tools.

\end{abstract}
\maketitle
\section{Introduction}\label{sec:introduction}

The growing complexity and scale of software systems have led to an increase in vulnerabilities~\cite{anwar2021cleaning}.
Recent advances in vulnerability detection tools have contributed to the identification of more vulnerabilities~\cite{10.1145/3699711}.
According to statistics from the National Vulnerability Database (NVD), a total of 40,009 software vulnerabilities were publicly disclosed in 2024, with a year-on-year increase of 38.83\%~\cite{nvd}.
Untreated vulnerabilities pose severe risks, including financial loss, data breaches, and systemic failures.
However, vulnerability repair remains labor-intensive, requiring security experts to analyze root causes, validate fixes, and ensure codebase compatibility, often consuming hundreds of developer hours per critical flaw~\cite{bitsight2024kev}.
Consequently, there is an urgent need for automated vulnerability repair (AVR) tools to reduce the cost.

\begin{figure}
    \centering
    \includegraphics[width=0.94\columnwidth]{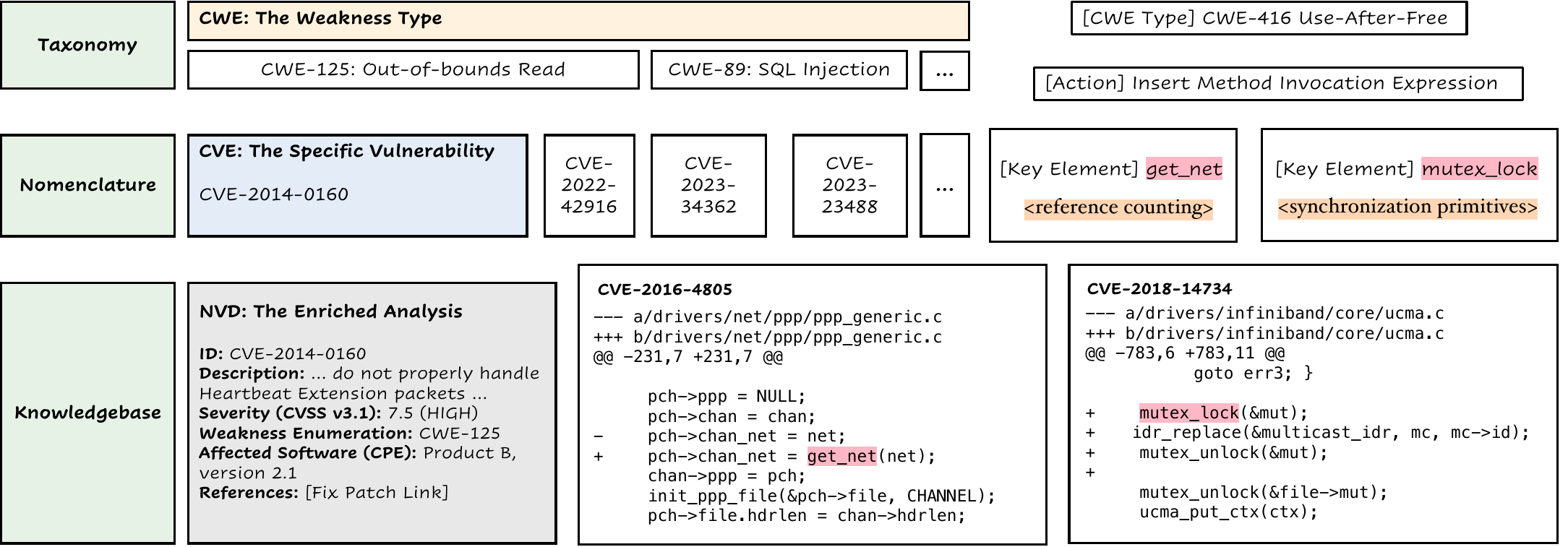}
    \caption{The Vulnerability Information Tracking Systems}
    \Description{A flow-style overview of vulnerability information tracking systems, showing the progression from CWE weakness categories to CVE vulnerability records and then to detailed NVD analyses and repair information.}
    \label{fig:track}
\end{figure}

Existing vulnerability repair studies can be broadly classified into two categories, \ie \textit{traditional program analysis-based approaches}~\cite{10.5555/2818754.2818812,10.1145/3236024.3236079,10.1145/3180155.3180250,10.1145/3377811.3380323} and \textit{learning-based approaches}~\cite{chi2020seqtrans,fu2022vulrepair,10.1145/3597503.3639222,vrepair}.
Traditional approaches rely on static or dynamic analysis with predefined rules, which are constrained by a finite modeling space.
This limited space restricts their ability to generalize to a wide range of vulnerabilities, often requiring manual tuning or specific rules to handle complex cases.
On the other hand, learning-based approaches, particularly those leveraging large language models (LLMs), offer a higher degree of generality and repair capability.
The advent of LLMs has significantly enhanced the effectiveness of large code models (LCMs), making them highly beneficial for AVR~\cite{10.1145/3597503.3639222}.
Recent state-of-the-art approaches in AVR~\cite{fu2022vulrepair,10.1145/3597503.3639222}, predominantly based on LCMs, have achieved remarkable improvements in repair performance, surpassing traditional approaches.

However, LLMs face a fundamental challenge in AVR. Their general-purpose code completion training is not inherently aligned with the specific requirements of security-driven patch generation. Although specialized security domain knowledge is crucial for guiding LLMs to generate secure and targeted fixes, current LLM-based repair approaches struggle to effectively integrate and apply this knowledge.
Particularly, this domain knowledge has been meticulously accumulated and formalized by vulnerability tracking systems, such as the Common Weakness Enumeration (CWE)~\cite{CWE}, Common Vulnerabilities and Exposures (CVE)~\cite{CVE}, and the National Vulnerability Database (NVD)~\cite{nvd}. 
As shown in Figure~\ref{fig:track}, the tracking systems systematically progress from abstract weakness classification (CWE) to specific identification (CVE) and detailed analysis, including repair patches (NVD).
This systematic classification and detailed documentation of vulnerabilities and their corresponding patches enable the extraction of invaluable, type-specific repair strategies (\eg employing reference counting or synchronization primitives for CWE-416 Use-After-free).

Despite the great potential of such structured, security-specific knowledge, existing LLM-based vulnerability repair approaches fall short in effectively leveraging it. 
For instance, VulRepair~\cite{fu2022vulrepair} and VRepair~\cite{vrepair} merely concatenate the CWE type to the vulnerable function as input without explicitly distinguishing between type-specific repair strategies, which yields negligible effects on repair accuracy (Table~\ref{tab:pre-study}). This is because the paradigm does not define a clear, strategy-grounded learning objective; instead, it expects the generator to implicitly learn the mapping from a coarse type label to the correct repair strategy, which often fails. VulMaster~\cite{10.1145/3597503.3639222} attempts to inject domain knowledge through concrete examples retrieved from the CWE website by adding a fixed set of demonstrations per CWE into the prompt.
However, our analysis indicates that VulMaster struggles to generalize beyond these fixed cases, limiting its effectiveness for vulnerabilities that deviate from textbook examples and require dynamic, context-aware solutions. Moreover, our exemplar-based few-shot/retrieval-augmented experiments in Section~\ref{sec:motivation} show that, unlike the short and curated CWE demonstrations, retrieving full real-world patch exemplars can degrade patch-generation performance because project-specific identifiers and incidental logic introduce contextual noise and induce superficial mimicry. These observations suggest that effectively leveraging vulnerability knowledge requires two ingredients: a compact, actionable representation beyond a coarse type label or noisy exemplars, and a learnable alignment/selection mechanism that chooses the right strategy for each vulnerability.

In the related field of Automated Program Repair (APR), instead of using demonstrative examples, traditional template-based approaches~\cite{10.1145/3510003.3510147,ntr,Liu_2019} summarize repair strategies as programmatically derived syntax actions (\eg Insert Missed Statement), and incorporate them into the prompt to guide the generation.
This abstraction helps mitigate the noise present in concrete examples and facilitates the identification of common repair strategies from diverse repair instances. However, these approaches are limited by their fixed pattern spaces, leading to template coverage problems; more fundamentally, a purely syntax-action abstraction is often overly generic: the same syntactic action may correspond to multiple security mechanisms, losing the important security semantics needed to characterize and guide vulnerability-specific fixes, as studied in Section~\ref{sec:motivation}.

To address these limitations, we propose \vulkey, an LLM-based AVR approach that leverages hierarchical abstraction of vulnerability expert knowledge to guide the model's repair code generation process.
Specifically, we summarize the CWE type-specific repair strategies into automatically generated repair patterns.
To capture essential repair details while maintaining generality, the repair patterns in \vulkey are formulated as a three-level abstraction, encompassing \textit{CWE type}, \textit{syntactic actions}, and \textit{semantic key elements}. Figure~\ref{fig:track} presents an example where \texttt{(CWE-416, Insert Method Invocation Expression, get\_net)} exemplifies a security repair strategy using reference counting, and \texttt{(CWE-416, Insert Method Invocation Expression, mutex\_lock)} exemplifies one using synchronization primitives. Here, a syntactic action denotes the core security-focused edit operator, such as inserting a missing check or mutating a condition. A semantic key element is a short code snippet extracted from the patch diff that instantiates the chosen action by explicitly specifying the security primitive or constraint to introduce.
Beyond forming a structured representation, we introduce a learnable alignment stage. A dedicated matcher selects the most suitable (Action, Key Element) pattern for each new vulnerability from this compact, discrete space, and we inject the selected pattern into the prompt as guidance during patch generation.
Compared to direct CWE-ID concatenation, the repair pattern explicitly instructs the model to follow the type-specific repair strategies by incorporating the repair patterns into the prompt, thereby guiding the search direction of LCMs toward effective vulnerability repair. 
Compared to concrete examples, the repair pattern captures the essential parts of the repair in a precise and concise manner, exhibiting stronger generality and facilitating better integration with the context.
Compared to syntax-based templates, our method generates repair patterns automatically summarized by the LLM, thereby addressing the template coverage problem caused by the limited scope of predefined templates. Furthermore, our method encodes semantic information into patterns to capture the underlying security principle.
This learnable alignment stage also brings three key advantages that complement our pattern formulation. First, we explicitly introduce a pattern-matching model (matcher) to enable context-aware pattern retrieval/strategy selection, providing a customized repair strategy for each target vulnerability. Second, because the matcher selects from a compact, discrete three-level pattern space, the prediction space is substantially narrowed, which improves selection accuracy. Third, since we inject the selected pattern only at inference time, alignment/selection is decoupled from patch generation, allowing the two stages to be optimized independently.

We design a two-stage pipeline to build \vulkey, consisting of an \textit{expert knowledge matching} stage and a \textit{repair code generation} stage.
In the first stage, we extract repair patterns from historical vul-fix diffs of 3,789 real-world vulnerabilities~\cite{primevul} and train a context-aware CodeT5p~\cite{codet5p2023}-based matcher. Given the CWE type and vulnerable-code context, the matcher predicts the top-$k$ most suitable (Action, Key Element) patterns from the discrete pattern space.
In the second stage, \vulkey undergoes progressive fine-tuning via transfer learning, leveraging both general bug-fix and security-specific vulnerability-fix corpora: it first learns general repair patterns from the extensive bug-fix data, and then uses the limited vulnerability-fix corpora to focus on security-specific repairs.
Finally, the top-$k$ (Action, Key Element) patterns selected in the previous stage are injected into the generation prompt at inference time as explicit guidance, enabling more precise vulnerability-specific repair generation.

\begin{table}
\centering
\caption{Performance of Concatenating CWE Type as Input Prefix. EM = Exact Match.}
\scriptsize
\begin{tabular}{lcc}
\toprule
\textbf{Implementation} & \textbf{StarCoderBase-15B (EM)} & \textbf{Qwen2.5-Coder-32B (EM)} \\
\midrule
Base finetune & 23.9 & 23.2 \\
\makecell{Base finetune with CWE type prefix (both train and inference)} & 23.6 & 22.5 \\
\bottomrule
\end{tabular}
\label{tab:pre-study}
\end{table}

We evaluate \vulkey on a real-world C/C++ vulnerability dataset \textit{PrimeVul}~\cite{primevul}, which is an enhanced version of widely-used benchmarks \textit{CVEFixes}~\cite{Bhandari_2021} and \textit{BigVul}~\cite{10.1145/3379597.3387501}. 
Experimental results demonstrate that \vulkey achieves state-of-the-art performance with 31.5\% repair accuracy, representing a 7.6\% absolute improvement over the best baseline StarCoder~\cite{starcoder2023} fine-tuned with bug-fix data and vulnerability-fix data.
It also significantly outperforms both task-specific approaches \textsc{VulRepair}~\cite{fu2022vulrepair} by 8.5$\times$, \textsc{VulMaster}~\cite{10.1145/3597503.3639222} by 3.6$\times$ and closed-source LLM \textsc{GPT-5} by 3$\times$.
To assess cross-language and cross-model generalizability, we evaluate \vulkey on the Java benchmark Vul4J.
\vulkey achieves 18 successful repairs in Vul4J, surpassing the state-of-the-art approaches VulMaster~\cite{10.1145/3597503.3639222} and VulRepair~\cite{fu2022vulrepair} (with 9 and 4 successful repairs, respectively).


This work makes the following major contributions:

\begin{itemize}[noitemsep,leftmargin=5.5mm]
    \item Existing LLM-based AVR approaches suffer from a critical limitation, \ie their insufficient integration of vital vulnerability domain knowledge.
    To address this, we propose a novel three-level abstraction for CWE expert knowledge, which formulates precise, type-specific repair patterns by capturing both syntactic actions and semantic key elements. This approach automatically summarizes flexible patterns, thereby overcoming the limitations of rigid, predefined templates and their inherent narrow coverage in traditional methods.

    \item We introduce \vulkey, a novel LLM-based AVR approach built upon a two-stage pipeline. This pipeline integrates a context-aware pattern predictor for expert knowledge matching with a progressively fine-tuned code generation model, effectively aligning LLMs for security-specific repair generation.

    \item We extensively evaluate \vulkey on the real-world C/C++ vulnerability dataset \textit{PrimeVul} and the Java benchmark \textit{Vul4J}. Our results demonstrate that \vulkey significantly improves repair effectiveness, achieving state-of-the-art performance across both benchmarks.
\end{itemize}

\section{Background}  

\subsection{Vulnerability Tracking Systems}
This process of leveraging domain knowledge is supported by a systematic information tracking system, as illustrated in Figure~\ref{fig:track}. The pipeline begins at the highest level of abstraction with the Taxonomy layer, where a weakness is categorized by its CWE~\cite{CWE} type (\eg CWE-416 Use-After-Free). From this general classification, the system moves to the Nomenclature layer, where a specific, real-world instance of that weakness is assigned a unique CVE~\cite{CVE} identifier (\eg CVE-2014-0160).
Finally, the Knowledgebase layer, often represented by resources like the NVD~\cite{nvd}, provides enriched, actionable details. This includes severity scores, affected software versions, and, most critically for automated repair, links to the exact code patches that fix the vulnerability. These patches represent concrete, expert-vetted solutions.
This rich, hierarchical structure provides a goldmine of information for learning repair strategies. For a single CWE type like CWE-416, the associated patches in the knowledge base reveal multiple distinct, yet effective, remediation patterns. For instance, as shown in Figure~\ref{fig:track}, one repair for CWE-416 might involve reference counting (abstracted from the key element get\_net), while another might use synchronization primitives (abstracted from mutex\_lock). Both repairs may share the same high-level syntactic action, such as \texttt{Insert Method Invocation Expression}, but present fundamentally different underlying security mechanisms. Capturing this nuance is essential for generating correct and context-aware patches, forming the foundational motivation for our work.

\subsection{Automated Vulnerability Repair}

AVR~\cite{zhou2025large} is a critical area that focuses on automatically fixing security flaws in software by patching. By addressing these flaws, AVR prevents their exploitation, thereby mitigating severe consequences such as data breaches, unauthorized access, and system disruption.
AVR shares the overarching goal of fixing software defects with APR~\cite{10.1145/3696450}. However, AVR presents distinct and often more challenging issues. Although existing approaches developed for general program repair might seem directly applicable to vulnerabilities, this is often not the case due to the unique characteristics of security flaws.

First, the nature of flaws differs significantly. While APR primarily addresses functional errors with localized fixes guided by failing tests, vulnerability repair often involves subtle, architectural weaknesses that demand an understanding of broader security design principles. Consider the Time-of-Check to Time-of-Use (TOCTOU) race condition vulnerability as an illustrative example:

\begin{itemize}[leftmargin=5.5mm, labelsep=0.5em, itemsep=0.5ex, topsep=0.5ex, parsep=0ex]
    \item \texttt{Check: if (access("user\_file.txt", W\_OK) == 0)}
    \item \texttt{Use: fd = open("user\_file.txt", O\_WRONLY)}
\end{itemize}

While the code lines are individually correct, the vulnerability lies in the time gap between the permission check and the subsequent file use. An attacker can exploit this by swapping the file with a symbolic link to a sensitive system file after the check but before its use. Fixing it is often indirect and requires security design knowledge (\eg atomic operations, principle of least privilege), typically by replacing the check-then-use pattern with safer OS-level primitives (\eg \texttt{seteuid}). This illustrates that vulnerability repair is fundamentally about security-oriented redesign rather than simple logical bug fixing.

Second, this distinction creates a profound gap in the repair and validation process. While APR relies on comprehensive test suites providing clear pass/fail feedback, vulnerability repair often lacks a pre-existing failing test; the test case is typically a complex, environment-dependent exploit. More critically, a successful patch must eliminate the entire class of underlying weakness without introducing new regressions, a much higher standard that simple tests cannot guarantee. This absence of a well-defined, binary validation oracle makes automated patch generation and verification exceptionally difficult.
Consequently, these fundamental challenges make feedback-driven, iterative APR methods, such as ChatRepair~\cite{chatrepair} and ThinkRepair~\cite{thinkrepair}, impractical for vulnerability repair. These methods depend on a tight loop of patch generation and simple test feedback, a paradigm that breaks down in the intricate and nuanced realities of security hardening.
Therefore, AVR requires a fundamental shift towards systems capable of reasoning about security principles and attacker behavior, which can generate correct patches based on security domain knowledge even in the absence of an oracle. The well-maintained vulnerability tracking systems, along with their accumulated data over time, provide the necessary foundation for this approach. Specifically, the vulnerability reports sourced from tracking systems provide structured data, including detailed descriptions, corresponding repair patches, and crucial classification by CWE ID. These data offer invaluable guidance for categorizing flaws and developing effective security-focused repair strategies.
\section{Preliminary Study and Motivation}\label{sec:motivation}  

Repairing vulnerabilities within the same CWE category often involves addressing similar underlying issues, as vulnerabilities grouped under a specific CWE typically share common root causes and manifestations~\cite{CWE}. For example, the CWE website explicitly describes recurring patterns and recommended mitigation strategies for each category, inherently suggesting the possibility of abstracting generic repair strategies for each CWE category.
To leverage this similarity, existing methods propose using few-shot in-context learning~\cite{Brown2020} to learn from concrete repair examples within the same CWE category. Our preliminary experiments demonstrate the ineffectiveness of concrete examples as repair strategies and further explore alternative representations.



\subsection{Concrete Example as Repair Strategy}

\begin{figure}
    \centering
    \begin{minipage}[b]{0.58\columnwidth}
        \centering
        \includegraphics[width=0.9\linewidth]{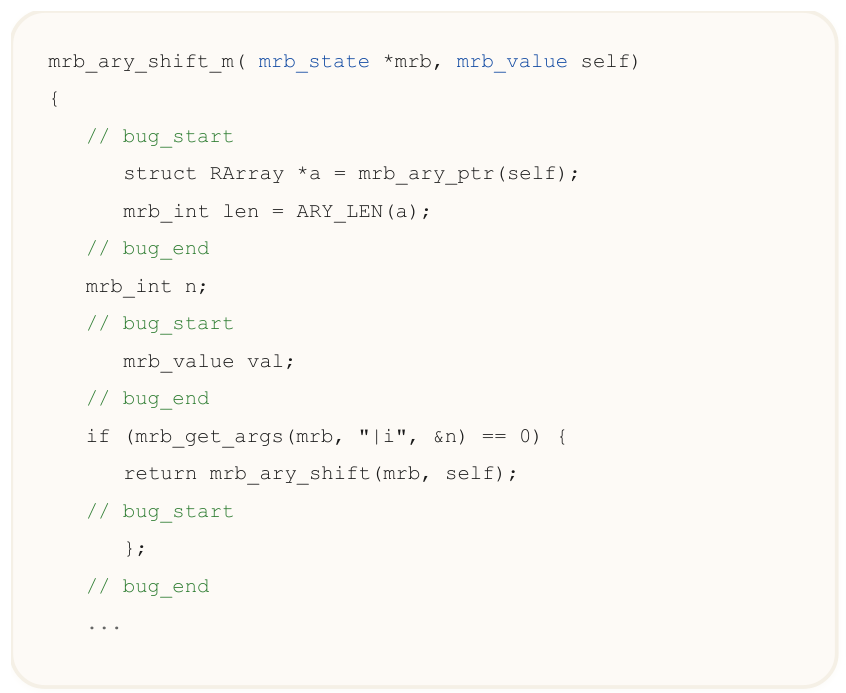}
        \caption*{(a) Bug/Vul Function}
        \label{fig:pair1}
    \end{minipage}
    \hfill
    \begin{minipage}[b]{0.4\columnwidth}
        \centering
        \includegraphics[width=\linewidth]{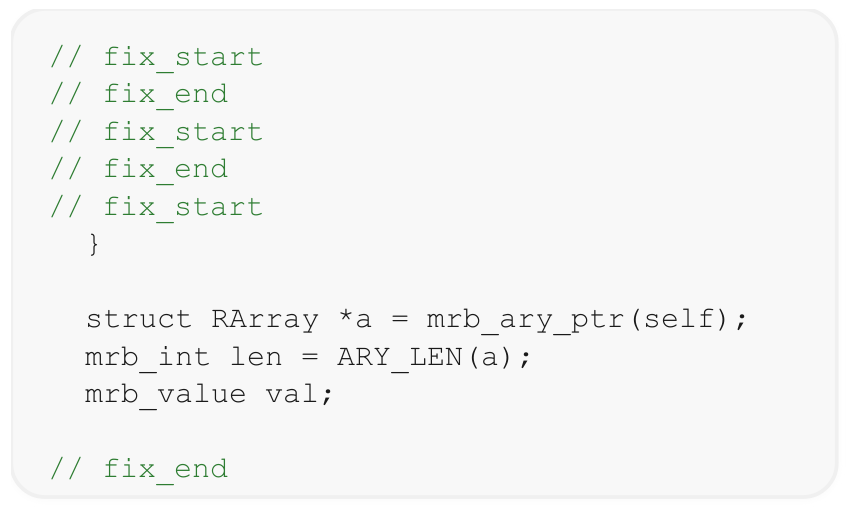}
        \caption*{(b) Fix Code}
        \label{fig:pair2}
    \end{minipage}
    \caption{Bug/Vul-fix Pair in \vulkey}
    \Description{Two side-by-side code examples: the left panel shows a buggy or vulnerable function with marked bug boundaries, and the right panel shows the corresponding fixed code extracted from the historical patch.}
    \label{fig:pair}
\end{figure}

To begin our investigation, we conducted preliminary experiments using the simplest form of repair strategy representation, \ie concrete examples. In this context, a concrete example refers to an actual repair patch, specifically represented as a pair consisting of a vulnerable function and its corresponding fixed version, as demonstrated in Figure~\ref{fig:pair}. We incorporated these concrete examples into the prompt to enable LLMs to perform in-context few-shot learning~\cite{Brown2020}. This approach allows us to directly leverage real-world repair instances as guidance for AVR. To evaluate the effectiveness of this representation, we study two strategies for the selection of concrete examples:

\begin{itemize}[noitemsep,leftmargin=5.5mm]
    \item \textit{Random Selection}: For each sample, we randomly selected as many vul-fix pairs with the same CWE type as possible (within the context window limit) to serve as concrete examples.

    \item \textit{Similarity-based Retrieval-Augmented Generation (RAG)}~\cite{fan2024survey}: For each sample, we ranked candidate vul-fix pairs within the same CWE type by lexical code similarity (BM25~\cite{robertson2009bm25}) between the query vulnerable code segment and each candidate's vulnerable code segment. We then selected as many top-ranked vul-fix pairs as possible (within the context window limit) to serve as concrete examples.
\end{itemize}

We evaluate 435 vulnerability instances on PrimeVul~\cite{primevul}, an upgraded version of existing real-world C/C++ vulnerability datasets. 
The metric we use is Exact Match (EM), which is a binary measure of repair effectiveness. Specifically, EM is calculated by determining if at least one of the 100 generated patches for an instance precisely matches the ground truth (after removing spaces and comments). The final EM score is then presented as the percentage of instances where such a match occurred.
The results of the study are presented in Table~\ref{tab:pre-study2}, which reveal that the indiscriminate use of concrete examples can lead to a significant performance degradation. In contrast, RAG based on code similarity can mitigate this issue, indicating that the accurate selection of domain knowledge can improve performance.

Nevertheless, compared to the base fine-tuning approach, the performance of both methods still deteriorates. This suggests that representing domain knowledge solely as concrete examples is suboptimal. First, full code examples are laden with context-specific details, such as variable names, custom data types, and logic that is peripheral to the core repair strategy. This extraneous information can act as noise within the context window, potentially distracting the model and hindering its ability to focus on the essential repair logic applicable to the new vulnerability. Second, models may struggle to generalize the underlying abstract repair pattern from a few concrete instances. Instead of identifying the fundamental principle, the model might attempt to mimic the verbatim code patterns of the provided examples, which is often unsuitable for the target code's unique context.

Taken together, these findings demonstrate that for domain knowledge to be effectively integrated, both accurate selection and a concise abstract representation are essential. Our work achieves this by condensing repair strategies into abstract patterns and employing a selection model under the fine-tuned paradigm based on the CWE type and vulnerable code segment.

\begin{table}[t]
\centering
\caption{Performance of In-context few-shot Learning Using Concrete Example. EM = Exact Match.}
\footnotesize
\begin{tabular}{lcc}
\toprule
\textbf{Implementation} & \textbf{StarCoderBase-15B (EM)} & \textbf{Qwen2.5-Coder-32B (EM)} \\
\midrule
Base Finetune & 23.9 & 23.2 \\
Base Finetune + Random & 11.9 & 12.9 \\
Base Finetune + Similarity-based RAG & 17.9 & 16.8 \\
\bottomrule
\end{tabular}
\label{tab:pre-study2}
\end{table}

\subsection{Abstraction for Repair Strategy}

To further investigate the design of domain knowledge representation, we now focus on higher-level abstractions of repair strategies. Prior template-based work~\cite{ntr} in APR abstracts repair strategies as syntax actions (\ie template), such as \texttt{Insert Missed Statement}, \texttt{Mutate Conditional Expression}. They are integrated into the prompt to guide the patch generation process of LLMs. These actions are characterized by AST-based structure matching. While such syntax-based paradigms are convenient for automatically mapping patches to corresponding templates, these overly coarse-grained actions are inadequate for fully capturing the complexity of security-specific repair strategies.
To convince that more fine-grained information is needed, we conducted a manual inspection of 416 repair instances within two common CWE categories in \textit{PrimeVul}~\cite{primevul}, an upgraded version of existing real-world C/C++ vulnerability datasets, utilizing the key elements extracted in Section~\ref{sec:extraction} for our analysis.

Our analysis of 249 repair instances for CWE-787 Out-of-bounds Write reveals that the most common repair strategies involve adding checks and constraints. Over half of the repairs (121 out of 234) fall under the syntactic template \texttt{Insert Range Checker}. However, this syntactic label is overly broad. The critical semantic information lies in the specific condition being checked, highlighted by our key elements. For instance, repairs include checking against a calculated length (\texttt{len}), a specific image dimension (\texttt{image\_height}), a hardcoded protocol limit (\texttt{if (payload\_size < 22) return;}), potential integer overflows (\texttt{*object + size < *object}), and array index boundaries (\texttt{(pps->sps\_id<0) || (pps->s\_id >= 16)}). Each key element represents a distinct semantic constraint required to fix the vulnerability, even though they all share the same syntactic template.


Our analysis of 167 repair instances for CWE-416 Use-After-Free (UAF) shows that repairs primarily focus on enforcing correct object lifecycle management and synchronization. The most frequent syntactic action is \textit{Insert Method Invocation Expression}, accounting for 27 repairs. Yet, this label alone fails to describe the repair's purpose. The key elements, however, reveal two dominant and distinct high-level strategies. Key elements like \texttt{mutex\_lock}, \texttt{snd\_use\_lock\_use}, and \texttt{mutex\_lock\_interruptible} represent the introduction of synchronization primitives. These patches fix race conditions where one thread might free an object while another is using it. In contrast, key elements such as \texttt{get\_net}, \texttt{get\_file\_rcu}, and \texttt{usb\_get\_intf} signify the implementation of reference counting. This strategy ensures an object's lifetime is extended by incrementing its usage count, preventing it from being deallocated prematurely. These two patterns, adding locks versus incrementing counters, are fundamentally different solutions to the UAF problem, a distinction completely lost without the semantic key elements.


This analysis underscores the critical role of semantic constraints in effectively addressing vulnerabilities. Thus, we argue that syntactic templates alone are insufficient to provide sufficiently precise guidance for remediation. To address this, our work organizes the pattern as multi-level representations that integrate both the syntax feature and the semantic feature. With this representation, our trained matching model identifies, for each vulnerable function, repair patterns that incorporate both syntactic and semantic information. These patterns, in turn, provide the patch generation model with more precise guidance for vulnerability remediation.


\section{Methodology}\label{sec:technique}

In this section, we introduce our proposed approach, \vulkey, which encodes domain knowledge as high-level abstracted repair patterns and employs a context-aware selection model to identify the most appropriate pattern to guide patch generation.
\vulkey also undergoes two-phase fine-tuning to progressively enhance learning of generic repair and security-specific repair.

\subsection{Overview}\label{sec:overview}

\begin{figure*}
    \centering
    \includegraphics[width=0.94\textwidth]{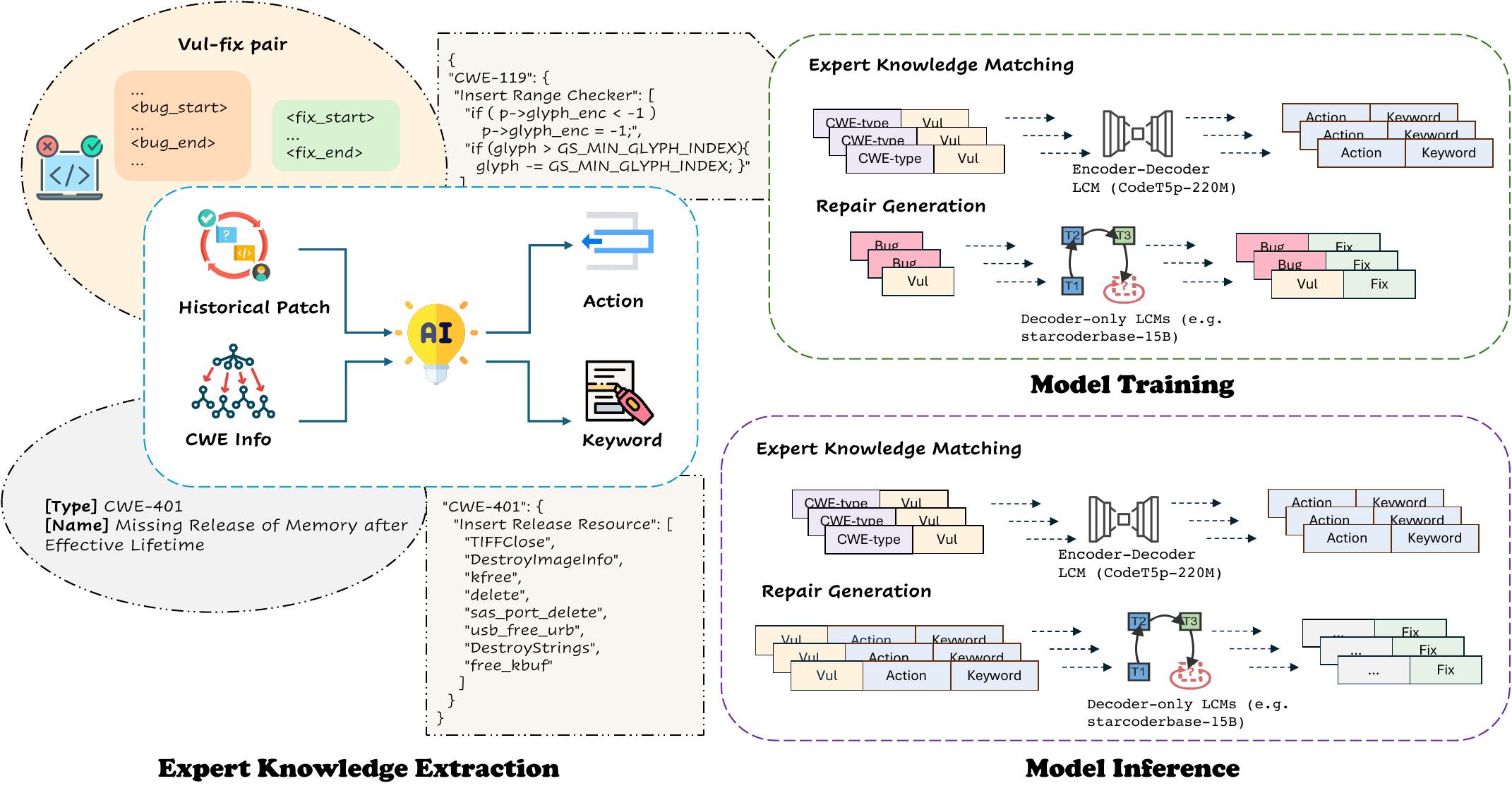}
    \caption{Overview of \vulkey}
    \Description{An overview of VulKey with three stages: expert knowledge extraction from historical vulnerability-fix pairs, model training for pattern matching and repair generation, and inference that first predicts action and key-element guidance and then generates repaired code.}
    \label{fig:arch}
\end{figure*}

\noindent\textbf{Inputs.} We assume three input signals, consistent with prior AVR studies~\cite{fu2022vulrepair,vrepair,10.1145/3597503.3639222,10.5555/3766078.3766309}. (1) A vulnerable function $X_i$ with the buggy region marked by \texttt{//bug\_start} and \texttt{//bug\_end}. (2) The CWE identifier $T_i$ and name $N_i$, produced by upstream detection/triage tools (\eg CodeQL~\cite{codeql}, SonarQube~\cite{sonarqube}); in our experiments we use dataset-provided labels. (3) Historical vul-fix pairs with ground-truth patches, code diffs $D_i$, vulnerability metadata (CVE descriptions, CWE type/name), and an initial action-option set for extracting repair patterns. We also use a large bug-fix corpus for two-phase fine-tuning of the repair generator.

As shown in Figure~\ref{fig:arch}, our framework consists of three components: \textit{expert knowledge extraction}, \textit{model training}, and \textit{model inference}. \textbf{Expert knowledge extraction} extracts actions $A_i$ and key elements $K_i$ from the code diff $D_i$ between the vulnerable function and its fixed version, based on the CWE type $T_i$ and its name $N_i$.
\textbf{Model training} involves two objectives, \ie matching action $A_i$ and key element $K_i$ to the vulnerable function $X_i$ and its CWE type $T_i$, and completing the buggy function $X_i$ with its repaired code $Y_i$.
\textbf{Model inference} uses a two-stage pipeline: first, it predicts actions $\hat{A_i}$ and key elements $\hat{K_i}$ given the vulnerable function $X_i$ and the CWE type $T_i$; then, it generates the repair code $\hat{Y_i}$ using $X_i$, $\hat{A_i}$, and $\hat{K_i}$. 
Notably, expert-knowledge extraction/matching is decoupled from patch generation. The repair generator is trained with a general completion objective $X_i \rightarrow (X_i, Y_i)$ without conditioning on Action or Key-Element inputs; at inference time, the matcher first predicts $\hat{A_i}$ and $\hat{K_i}$, which are then injected as guidance. This separation prevents co-adaptation and lets us improve or replace the action inventory, the matcher, or the underlying code-generation LLM without knowledge-specific retraining of the repair generator.

In the rest of this section, we elaborate on these components in sequence.

\subsection{Expert Knowledge Extraction}\label{sec:extraction}

Vulnerability tracking systems accumulate substantial expert repair knowledge across CWE types, largely embodied in security fix patches; however, this knowledge remains unstructured and dispersed in datasets. We extract key actionable information from patches and formalize it into reusable repair patterns. Unlike prior hand-labeled or programmatically derived patterns, we automate extraction using LLMs' code comprehension ability and distill each historical vul-fix pair into a compact (Action, Key Element) pair $(A_i, K_i)$.

First, $A_i$ is a discrete \textit{syntactic action} that names the syntactic edit operator embodying the core security repair principle of the patch, describing \textit{how} to edit the code. Examples include inserting a missing check, mutating a condition, and inserting a resource-release call. We start from a seed action inventory adopted from prior template-based APR works~\cite{10.1145/3510003.3510147,Liu_2019}, which covers common syntax-level edit operators, as listed in Table~\ref{tab:actions}. When the seed inventory cannot adequately describe a patch, we allow the LLM to propose a new action label. We then cluster newly proposed labels and merge semantically equivalent ones into canonical actions with brief definitions, keeping the action space compact yet extensible.

Second, given a selected action $A_i$, the Semantic Key Element $K_i$ is a short, high-signal contiguous snippet, typically an expression or a simple statement, extracted from the added \texttt{+} lines of a security patch diff. It captures the core security mechanism or constraint introduced by the fix, \ie what to introduce, rather than project-specific implementation details or incidental, non-security logic. Since $K_i$ semantically instantiates $A_i$, it disambiguates strategies that share the same action. For checker-type actions, \eg \texttt{Insert Null Pointer Checker}, \texttt{Insert Range Checker}, and \texttt{Insert Conditional Expression}, $K_i$ is usually the inserted guard predicate, such as \texttt{input == nullptr}, \texttt{remain<5}, or \texttt{ret < 0}. For API/primitive insertion actions, \eg \texttt{Insert Method Invocation Expression} and \texttt{Insert Release Resource}, $K_i$ is the inserted call expression, such as \texttt{gf\_bs\_align(bs)}, \texttt{kfree(a)}, or \texttt{usb\_put\_dev(udev)}. For mutation-type actions, $K_i$ corresponds to the modified critical fragment that enforces the intended constraint, such as \texttt{val \&\& len > 0}, \texttt{tmplen < 64}, or \texttt{crypto\_memneq(md, tmp, len)}.
To keep $K_i$ high-signal and consistent, we apply three rules. First, \textit{Semantic focus}: prioritize fragments that most directly embody the security strategy (\eg security-critical API/primitive calls, core guard predicates, and security-relevant types or flags) and avoid dilution by project-specific implementation details. Second, \textit{Minimal self-containedness}: select the smallest fragment that independently expresses the security mechanism (typically an expression or a simple statement) and avoid concatenating unrelated statements or entire code blocks. Third, \textit{Denoising}: filter out low-signal content such as logging, formatting, and temporary variables.

In implementation, for each vul-fix pair, we compute a line-level code diff $D_i$ using the Python library \textit{difflib} and collect vulnerability metadata, including CWE type/name and the CVE description. We then prompt the LLM GPT-4.1~\cite{openai} with the diff, metadata, and an initial action-option set to output an \texttt{action:key\_element} pair. To make $K_i$ checkable and auditable, we enforce programmatic substring validation: $K_i$ must be a contiguous snippet from the added lines (the sole exception is \textit{Remove Buggy Statement}, whose $K_i$ is drawn from the deleted lines). If validation fails, we re-run extraction up to three times and discard the pattern only if all attempts fail. The prompt is:

\begin{center}
\begin{minipage}{0.96\columnwidth}
\begin{lstlisting}[basicstyle=\tiny\ttfamily, columns=fullflexible, breaklines=true, breakatwhitespace=true, aboveskip=6pt, belowskip=6pt, breakindent=0pt, frame=single, rulecolor=\color{black}, xleftmargin=2pt, xrightmargin=2pt, framesep=2pt]
{diff}
{cve_desc}
You are a security vulnerability repair expert reviewing the patch commit for {cwe} {cwe_name}.
Analyze the patch diff and extract ONE security repair pattern.

Step 1 (Select Action): choose ONE action from: {list(repair_actions)} that best describes the syntactic edit operator embodying the core security repair principle of this patch; if none fits, create a new action label in the same style.

Step 2 (Extract Key Element): given the selected action, choose ONE key element from the added lines (``+'' lines) that semantically instantiates the action (for Remove Buggy Statement, choose from the deleted lines (``-'' lines) instead).
- It MUST be copied verbatim from the added code and be a short contiguous snippet.
- Pick the smallest self-contained fragment that best captures the security mechanism/constraint introduced by the fix; e.g., a guard predicate or a security-critical call.
- Avoid low-signal noise, e.g., large blocks, logging, comments, and temporary variables, and favor security-relevant calls/checks/types over incidental implementation details.

Output (STRICT): output EXACTLY ONE line in the form: action:key_element (no extra text).

Example output: {Insert Release Resource:delete}
\end{lstlisting}
\end{minipage}
\end{center}

Ultimately, we derive combinations of these generated actions and key elements, which serve to characterize the repair pattern for each vulnerability-fix pair. These are then organized into a three-level expert knowledge representation $(X_i, T_i, A_i, K_i)$, with each entry representing a repair pattern classified by CWE type and vulnerable function.

\begin{table}[t]
\centering
\caption{List of Repair Actions (Entries not included in the initial set are \colorbox{gray!25}{grayed out})}
\label{tab:actions}
\scriptsize
\setlength{\tabcolsep}{4pt}
\begin{tabular}{clclcl}
\toprule
\textbf{No.} & \textbf{Repair Actions} & \textbf{No.} & \textbf{Repair Actions} & \textbf{No.} & \textbf{Repair Actions} \\
\midrule
1  & \cellcolor{gray!25}Insert Variable             & 7  & Insert Null Pointer Checker         & 13 & Mutate Literal Expression           \\
2  & \cellcolor{gray!25}Insert Memset               & 8  & Insert Missed Statement             & 14 & Mutate Method Invocation Expression \\
3  & \cellcolor{gray!25}Insert Release Resource     & 9  & \cellcolor{gray!25}Mutate Control Statement & 15 & Mutate Return Statement             \\
4  & \cellcolor{gray!25}Insert Cast Statement       & 10 & \cellcolor{gray!25}Insert Conditional Expression & 16 & Mutate Variable                     \\
5  & Insert Cast Checker             & 11 & Mutate Conditional Expression       & 17 & Move Statement                      \\
6  & Insert Range Checker            & 12 & \cellcolor{gray!25}Insert Method Invocation Expression & 18 & Remove Buggy Statement              \\
\bottomrule
\end{tabular}
\end{table}

\subsection{Model Training}\label{sec:train}

\noindent\textbf{\textit{Bug/Vul-Fix Pairs.}}
Before training, we prepare our training data in the form of function-level bug/vulnerability-fix pairs, extracted from historical fix patches. Figure~\ref{fig:pair} shows an example.
In subfigure (a), we present a function that contains a bug or vulnerability. This function is extracted from a historical commit where the issue was identified and subsequently fixed. Within this buggy/vulnerable function, we delineate the changed code sections using \textit{//bug\_start} and \textit{//bug\_end} comments.
Subfigure (b) displays the corresponding fix for the function shown in subfigure (a). This fix is also extracted from the historical commit, representing the corrected version of the function after the bug or vulnerability has been addressed. The fix code includes the corresponding modifications for the bug/vulnerability section within the buggy/vulnerable function, which are similarly surrounded by \textit{//fix\_start} and \textit{//fix\_end} comments. It is important to note that our approach supports multi-hunk fixes.
By pairing the buggy or vulnerable function $X_i$ with its fixed code $Y_i$, we construct a training dataset $(T_i, X_i, Y_i, A_i, K_i)$ from which the model can learn.

\noindent\textbf{\textit{Expert Knowledge Matching.}}
After obtaining the training dataset, we start training the expert knowledge matching model, which will later be used to produce the repair pattern to guide the code generation in the inference phase.
As shown in Figure~\ref{fig:arch}, we formulate this problem as a sequence-to-sequence task: \((T_i, X_i) \rightarrow (A_i, K_i)\). Specifically, the input to the model is the concatenation of the CWE type \(T_i\) and the vulnerable code function \(X_i\), denoted as \(T_i \oplus X_i\). The output is the concatenation of the action sequence \(A_i\) and the key element sequence \(K_i\), denoted as \(A_i \oplus K_i\).

We select encoder-decoder LCM CodeT5p~\cite{codet5p2023} as the foundation model. In this context, the fine-tuning objective for the expert knowledge matching model \(\mathcal{M}_{match}\) is to minimize the cross-entropy loss \(\mathcal{L}_{match}\), thereby optimizing the model’s parameters \(\theta\) by reducing the discrepancy between the predicted sequences \(\hat{A}_i \oplus \hat{K}_i\) and the actual sequences \(A_i \oplus K_i\) for all concatenated inputs \(T_i \oplus X_i\). This objective is mathematically formulated as:

{\small
\begin{equation}
\setlength{\abovedisplayskip}{3pt}
\setlength{\belowdisplayskip}{3pt}
\min_{\theta} \mathcal{L}_{\text{match}} = - \sum_{i=1}^{N} \sum_{t=1}^{|A_i \oplus K_i|} \log P\left( (A_i \oplus K_i)_t \big| T_i \oplus X_i, (A_i \oplus K_i)_{<t}; \theta \right)
\end{equation}
}
where $\oplus$ denotes sequence concatenation, $N$ the training set size, and $|A_i \oplus K_i|$ the token count of the target sequences. The notation \((A_i \oplus K_i)_t\) refers to the (t)-th ground-truth token in the concatenated sequence, with \(  (A_i \oplus K_i)_{<t}\) representing the preceding gold tokens. The probability \(P\left( (A_i \oplus K_i)_t \big| T_i \oplus X_i, (A_i \oplus K_i)_{<t}; \theta \right)\) computes the likelihood of generating the target token conditioned on the vulnerability context and preceding oracle tokens, parameterized by \(\theta\).
By minimizing this cross-entropy loss, the model learns to generate precise and contextually relevant action and key element sequences given the CWE type and vulnerable function, which is essential for guiding the subsequent code generation process and enhancing the effectiveness of vulnerability repairs.

The reason why we choose encoder-decoder LCM CodeT5p~\cite{codet5p2023} lies in the fact that vulnerability repair not only demands an understanding of the semantics of the vulnerable code and CWE type, but also the generation of repair patterns to guide the repair generation. The encoder-decoder architecture of CodeT5p is well-suited for handling the mapping from code to natural language, making it an effective choice for this task. In contrast, prior studies have demonstrated that other LCMs, such as encoder-only models (\eg CodeBERT~\cite{feng2020codebert}) and decoder-only models (\eg CodeLlama~\cite{roziere2023codellama}), exhibit suboptimal performance and efficiency in the code-to-natural-language matching task~\cite{10.1145/3691620.3695555}.

\noindent\textbf{\textit{Repair Generation.}} To guide the direction of LCM code generation, we employ a two-phase fine-tuning process under transfer learning. This approach leverages the extensive availability of bug-fix datasets as a supplement to the relatively limited vul-fix datasets. Initially, we fine-tune the model on the large bug-fix dataset to exploit the abundance of data, facilitating effective model convergence and enabling the model to learn general patterns and structures related to code fixes. Subsequently, fine-tuning on the smaller vul-fix dataset allows the model to specialize in vulnerability repair, thereby enhancing its effectiveness in generating security-specific fixes.
Given the bug/vulnerable function \(X_i\), we formulate repair generation as a completion task \(X_i \rightarrow (X_i, Y_i)\). The objective of the repair generation model, \(\mathcal{M}_{repair}\), is to auto-regressively synthesize the appropriate fix code \(\hat{Y}_i\) given \(X_i\). In practice, we use model-specific special tokens to separate the bug/vulnerable function and the fix code. 

We select decoder-only LCMs as the foundation model to benefit from their superior code generation ability. Given an input sequence \(X_i\) and a target sequence \(Y_i\), the loss function is formulated as:

{\small
\begin{equation}
\setlength{\abovedisplayskip}{1pt}
\setlength{\belowdisplayskip}{3pt}
\mathcal{L}_{\text{repair}} = - \sum_{i=1}^{N}\sum_{t=1}^{T} \log P\left( Y_{i,t} \big| X_i, Y_{i,<t} \right)
\end{equation}
}
where \(T\) denotes the length of the target sequence \(Y_i\), \(Y_{i,t}\) represents the \(t\)-th token of the target sequence, and \(Y_{i,<t}\) signifies all tokens in the target sequence preceding the \(t\)-th token. The term \(P(Y_{i,t} | X_i, Y_{i,<t})\) corresponds to the probability of predicting the target token \(Y_{i,t}\) given the input sequence \(X_i\) and the preceding tokens \(Y_{i,<t}\). The goal of fine-tuning is to maximize this function, thereby enhancing the model’s ability to accurately predict the next token in the sequence.

\subsection{Model Inference}
We define this stage as a pipeline consisting of expert knowledge matching \((T_i, X_i) \rightarrow (\hat{A_i}, \hat{K_i})\) and repair generation \((X_i, \hat{A_i}, \hat{K_i}) \rightarrow \hat{Y_i}\). The output of the first stage serves as the input for the second stage, ultimately achieving the object \((T_i, X_i) \rightarrow \hat{Y_i}\). 

\noindent\textbf{\textit{Expert Knowledge Matching.}} During inference, we employ beam search~\cite{Freitag_2017} with width $(B=10)$ to generate multiple candidate sequences of actions and key elements. This strategy enhances the diversity of guidance while maintaining the generation quality. The beam search objective for sequence generation is formulated as:

{\small
\begin{equation}
\setlength{\abovedisplayskip}{3pt}
\setlength{\belowdisplayskip}{3pt}
(\hat{A}_i, \hat{K}_i) = \underset{\substack{A^{(j)}, K^{(j)} \\ j \in [1,10]}}{\arg\max} \sum_{t=1}^{|A_i \oplus K_i|} \log P\left( (A_i \oplus K_i)_t \mid T_i \oplus X_i, (\hat{A}_i \oplus \hat{K}_i)_{<t} \right)
\end{equation}
}

The top-10 candidates \(\{(\hat{A}^{(j)}_i, \hat{K}^{(j)}_i)\}_{j=1}^{10}\)  provide directions for repair generation. Beam search outperforms stochastic sampling methods (nucleus/temperature) due to its ability to produce high-quality samples. This superiority is attributed to beam search's global sequence optimality achieved through joint probability maximization.

\noindent\textbf{\textit{Repair Generation.}} Upon acquiring the repair action $\hat{A}_i$ and key element $\hat{K}_i$, we systematically concatenate each action, key element pair with the corresponding vulnerable function $X_i$. This structured input is subsequently processed by the code repair generation model $\mathcal{M}_{repair}$ through the following formulation:

\begin{equation}
\setlength{\abovedisplayskip}{3pt}
\setlength{\belowdisplayskip}{3pt}
\hat{Y}_{ij} = \mathcal{M}_{repair}\left(X_i \oplus \hat{A}^{(j)}_i \oplus \hat{K}^{(j)}_i\right)
\end{equation}

where $\oplus$ denotes the concatenation operator. 
This dual-dimensional guidance mechanism (along both action and key element dimensions) effectively constrains the code generation search space within vulnerability repair, thus enhancing exploration efficiency.
To balance diversity and computational efficiency, we adopt temperature-controlled stochastic sampling during generation. This technique enables the model to produce multiple repair candidates through a single forward pass, achieving polynomial-time complexity $O(n)$ while maintaining sample variance.

\section{Experiment}\label{sec:evaluation}

To systematically investigate the effectiveness of our approach, we propose four questions and design experiments to answer them.
\begin{itemize}[leftmargin=*]

\item  \textbf{RQ1:} What is the overall performance of our \vulkey in repairing vulnerabilities compared to the baselines?
\item  \textbf{RQ2:} What is the quality of our extracted/matched expert knowledge patterns?
\item  \textbf{RQ3:} What is the contribution of each component in our approach?
\item  \textbf{RQ4:} What is the generalizability to another programming language?

\end{itemize}

\subsection{Dataset}\label{sec:dataset}
\noindent\textbf{\textit{Transfer Dataset.}} Transfer dataset~\cite{10.1145/3510003.3510147} comprises approximately one million bug-fix pairs of Java language. Given the substantial computational cost associated with training LLMs, we directly use the randomly selected 101,475 samples from this dataset by NTR~\cite{ntr} for our experiments. The distribution of these samples is as follows: Train/Validation/Test = 97,416/2,029/2,030. We utilize this dataset as the bug-fix training dataset for transfer learning of the repair generation model.

\noindent\textbf{\textit{PrimeVul Dataset.}} \textit{PrimeVul}~\cite{primevul} is an upgraded version of existing real-world C/C++ vulnerability datasets. It aggregates security-related commits from four established sources: BigVul~\cite{10.1145/3379597.3387501}, CrossVul~\cite{10.1145/3468264.3473122}, CVEfixes~\cite{Bhandari_2021}, and DiverseVul~\cite{10.1145/3607199.3607242}. Through rigorous deduplication and quality filtering, PrimeVul improves upon existing benchmarks by achieving: (1) higher label accuracy through commit-level verification and the only changed function filtering, (2) a rigorous data de-duplication and chronological data splitting strategy (3,789/480/435)  to mitigate data leakage issues. The final collection contains 5,480 vulnerability-fix pairs spanning 140 CWEs. This dataset has three roles: the vul-fix training dataset for transfer learning in the repair generation model, the training set of the expert knowledge matching model for C/C++, and the evaluation benchmark for vulnerability repair tasks for C/C++.

\noindent\textbf{\textit{X\textsubscript{1} Dataset.}} This Java-specific dataset~\cite{Shestov2024-finetunellmsvuldet} contains 1,334 carefully curated samples (810/272/252) focusing on single-function vulnerability fixes. By exclusively selecting commits modifying individual functions, it minimizes label ambiguity from unrelated code changes. We use this dataset to train the expert knowledge match model for the Java language. We confirmed no overlaps on \textit{Vul4J} test benchmarks to ensure evaluation validity.

\noindent\textbf{\textit{Vul4J Benchmark.}} Following~\cite{10.1145/3597926.3598135}, we evaluate on 35 single-hunk vulnerabilities from Vul4J~\cite{10.1145/3524842.3528482}. This benchmark provides 79 reproducible Java vulnerabilities from 51 OSS projects, each accompanied by ground-truth patches and PoV test cases.

\Needspace{5\baselineskip}  
\subsection{Baseline}\label{sec:baseline}
\noindent\textbf{\textit{Template-based APR approaches.}} We use recent template-based APR work NTR~\cite{ntr} as baseline.

\noindent\textbf{\textit{Learning-based AVR approaches.}} We use recent AVR works as baselines, including VulRepair~\cite{fu2022vulrepair}, and VulMaster~\cite{10.1145/3597503.3639222}.

\noindent\textbf{\textit{General-purpose large code models.}} We include decoder-only LCMs of different parameter sizes: CodeLlama-70B~\cite{codellama2024}, StarCoderBase-15B~\cite{starcoder2023}, DeepSeek-Coder-V2-Lite-Base-16B~\cite{dscoder2024}, and Qwen2.5-Coder-32B~\cite{qwencoder2024}. Among these, StarCoder and CodeLlama are well-established LLMs for code, while DeepSeek-Coder and Qwen2.5-Coder represent the latest advancements in code LLMs.

\noindent\textbf{\textit{Closed-source LLM.}} We include the well-known ChatGPT model (\ie gpt-5 and gpt-4.1) as the baseline~\cite{openai}. We utilize few-shot learning to teach LLM the output format. The prompt used is as follows, with the example at the end shown in Figure~\ref{fig:pair}:
\begin{center}
\begin{minipage}{0.96\columnwidth}
\begin{lstlisting}[basicstyle=\tiny\ttfamily, columns=fullflexible, breaklines=true, breakatwhitespace=true, aboveskip=3pt, belowskip=0pt, breakindent=0pt, frame=single, rulecolor=\color{black}, xleftmargin=2pt, xrightmargin=2pt, framesep=2pt]
{vulnerable_code}\n If you are a software engineer tasked with repairing vulnerabilities for {cwe_type} {cwe_name}, generate fixed code to substitute each code segment enclosed by // bug_start and // bug_end. You can delete, update, or insert code inside. Please limit your response to the fixed code of the vulnerable function exclusively. Here's an example: 
Input: {buggy_code_example}
Output: {fixed_code_example}
\end{lstlisting}
\end{minipage}
\end{center}

\subsection{Settings}\label{sec:metrics}
\noindent\textbf{\textit{Evaluation Metric.}} We use Exact Match (EM) to evaluate the effectiveness of the repair. For each repair instance, EM is a binary metric, yielding either true or false. Specifically, EM is considered true for a given sample if at least one of the generated patches has the exact same token sequence as the ground truth, after removing spaces and comments. In our experiments, the final EM score is presented as a percentage, which represents the proportion of all test set samples for which EM was evaluated as true.

\noindent\textbf{\textit{Expert Knowledge Matching.}} We use CodeT5p-220M~\cite{codet5p2023} as the base model for this stage due to its balanced capabilities in comprehension and generation, as well as its superior performance in classification tasks~\cite{10.1109/ICSE48619.2023.00180}. We trained the model for 10 epochs and selected the last checkpoint to ensure convergence and adequate training. For the training hyperparameters, we set the learning rate to 1e-5, the maximum input/output sequence length to 512. In the inference phase, we sample 10 guidance for every vulnerable function by beam search.

\noindent\textbf{\textit{Code Generation.}} We include two classic code LLMs (CodeLlama-70B~\cite{codellama2024}, StarCoderBase-15B~\cite{starcoder2023}) and two advanced code LLMs (DeepSeek-Coder-V2-Lite-Base-16B~\cite{dscoder2024}, Qwen2.5-Coder-32B~\cite{qwencoder2024}) to fully leverage the code comprehension and generation ability of code LLMs for repair generation. Our selection is diverse in parameter size to study the relation between parameter size and model ability. 
To reduce the GPU memory and computational consumption, we fine-tune these models using LoRA~\cite{hu2021loralowrankadaptationlarge} and Bit Quantization~\cite{quantization} (also called QLoRA~\cite{NEURIPS2023_1feb8787}). For training hyper-parameter settings, we set the learning rate to 5e-5 and the maximum input/output length to 2048; In the inference phase, we use temperature sampling, and for base fine-tuning, we sample 1/10/100 samples for 10 different temperatures, which comprise 10/100/1000 total sample size; for our approach \vulkey, we sample 1/10/100 samples for 10 guidance generated from previous expert knowledge matching phase with temperature=1.0, which also comprise 10/100/1000 total sample size.

\noindent\textbf{\textit{Environments.}} The implementation is based on Python 3.12 and PyTorch 2.5.1, with all models sourced from HuggingFace. The experiments are conducted on a 12-core workstation equipped with an Intel(R) Xeon(R) Platinum 8374C CPU, 2TB RAM, and 8 $\times$ 40G TESLA A100 GPUs, running Ubuntu 20.04.6 LTS.
 
\subsection{Experiment Results}

\begin{table}[t]
\scriptsize
\centering
\caption{Performance Comparison of Vulnerability Repair Approaches}
\label{tab:results}
\renewcommand{\arraystretch}{1.10}
\setlength{\tabcolsep}{6pt}
\begin{tabular}{llrr@{\hspace{6pt}}|@{\hspace{6pt}}llrrrr}
\toprule
\multirow{2}{*}{\textbf{Type}} & \multirow{2}{*}{\textbf{Approach}} & \multirow{2}{*}{\textbf{Sample}} & \multirow{2}{*}{\textbf{EM}} & \multirow{2}{*}{\textbf{Type}} & \multirow{2}{*}{\textbf{Model}} & \multicolumn{2}{c}{\textbf{Bug-fix}} & \multicolumn{2}{c}{\textbf{Vul-fix}} \\
\cmidrule(lr){7-8} \cmidrule(lr){9-10}
 & & & & & & \textbf{Sample} & \textbf{EM} & \textbf{Sample} & \textbf{EM} \\
\midrule
\multirow{3}{*}{\textbf{Task-specific}} & \textsc{VulRepair} & 100 & 3.7 & \multirow{4}{*}{\textbf{LCM}} & DeepSeekCoder & 100 & 17.0 & 100 & 15.2 \\
& \textsc{VulMaster} & 10 & 8.7 &  & CodeLlama & 10 & 17.0 & 10 & 20.4 \\
& \textsc{NTR} & 100 & 22.8 &  & QwenCoder & 10 & 20.6 & 10 & 23.6 \\
\cmidrule(lr){1-4}
\multirow{2}{*}{\textbf{Closed-source}} & GPT-5 & 10 & 10.6 &  & StarCoder & 100 & 20.4 & 100 & 23.9 \\
& GPT-4.1 & 10 & 10.1 &  &  &  &  &  &  \\
\cmidrule(lr){1-4} \cmidrule(lr){5-10}
\textbf{Ours} & $VulKey_{sc}$ & 100 & \textbf{31.5} & \textbf{Ours} & $VulKey_{sc}$ & \multicolumn{2}{c}{\textsc{n/a}} & 100 & \textbf{31.5} \\ 
\bottomrule
\end{tabular}
\end{table}

\subsubsection{RQ1: What is the overall performance of our \vulkey in repairing vulnerabilities compared to the baselines?}
\ 
\newline
\noindent\textbf{\textit{Setup.}} We compare \vulkey with the baselines on the \textit{PrimeVul} test set using Exact Match. All open-source LCMs are fine-tuned in two stages with bug-fix data \textit{Transfer} and vul-fix data \textit{PrimeVul}. VulRepair and VulMaster are retrained on preprocessed \textit{PrimeVul} data following their original settings~\cite{fu2022vulrepair,10.1145/3597503.3639222}, while GPT-5 and GPT-4.1 are queried via API with few-shot prompting. Based on the base-model results, we choose StarCoderBase-15B, which achieves the best performance among all base models, as the generation backbone of \vulkey, denoted as $VulKey_{sc}$; it is fine-tuned with bug-fix and vul-fix data and paired with an expert knowledge matcher trained on \textit{PrimeVul}. Following NTR~\cite{ntr}, we use sample size 10 for larger models and 100 for smaller ones to ensure comparable computational budgets across models; VulMaster is also limited to 10 due to its higher memory cost.

\noindent\textbf{\textit{Results.}} As evidenced in Table~\ref{tab:results}, \vulkey achieves state-of-the-art performance, \ie \textbf{31.5\%}, representing \textbf{7.6\%} improvement over the strongest baseline (StarCoder + bug-fix data + vul-fix data). Three key observations emerge: (1) Our two-phase fine-tuning strategy yields consistent gains across most architectures, with vul-fix data enhancing Codellama's EM by 20.0\% (17.0→20.4), QwenCoder's EM by 14.6\% (20.6→23.6), and StarCoder's EM by 17.1\% (20.4→23.9). (2) Architectural specialization matters: With almost equal parameter size and the same sample size (15B versus 16B), StarCoder-based implementations consistently outperform DeepSeekCoder-based implementations. Moreover, QwenCoder-based implementations consistently outperform Codellama-based implementations with a smaller parameter size (32B versus 70B).  (3) The expert knowledge integration in \vulkey delivers additional \textbf{32.0\%} relative improvement over standalone StarCoder with bug-fix and vul-fix data (23.9→31.5), significantly surpassing both task-specific (8.5$\times$ VulRepair), (3.6$\times$ VulMaster), (1.4$\times$ NTR) and closed-source LLM approaches (3$\times$ GPT-5).

\begin{table*}[htbp]
\centering
\caption{Vulnerability Repair Performance Comparison across Different CWE types. Success = Number of successfully repaired (exactly matching) cases. Rate = Success/Total $\times$ 100\% .}
\label{tab:cwe_results}
\scriptsize
\begin{tabular}{@{}llrrrrcrrr@{}}
\toprule
\multirow{2}{*}{\textbf{CWE ID}} & \multirow{2}{*}{\textbf{CWE Name}} & \multicolumn{3}{c}{\textbf{StarCoder+Bug+Vul}} & \phantom{abc} & \multicolumn{3}{c}{\vulkey} & \multirow{2}{*}{\textbf{Rate Change}} \\
\cmidrule(lr){3-5} \cmidrule(lr){7-9}
 & & \textbf{Success} & \textbf{Total} & \textbf{Rate (\%)} & & \textbf{Success} & \textbf{Total} & \textbf{Rate (\%)} &  \\
\midrule
CWE-787 & Out-of-bounds Write & 18 & 72 & 25.00 && 26 & 72 & 36.11 & $\uparrow$11.11\% \\
CWE-125 & Out-of-bounds Read & 11 & 48 & 22.92 && 14 & 48 & 29.17 & $\uparrow$6.25\% \\
CWE-476 & NULL Pointer Dereference & 11 & 39 & 28.21 && 14 & 39 & 35.90 & $\uparrow$7.69\% \\
CWE-416 & Use After Free & 3 & 29 & 10.34 && 6 & 29 & 20.69 & $\uparrow$10.34\% \\
CWE-200 & Sensitive Info Exposure & 6 & 16 & 37.50 && 7 & 16 & 43.75 & $\uparrow$6.25\% \\
CWE-20 & Improper Input Validation & 5 & 15 & 33.33 && 7 & 15 & 46.67 & $\uparrow$13.33\% \\
CWE-617 & Reachable Assertion & 4 & 12 & 33.33 && 6 & 12 & 50.00 & $\uparrow$16.67\% \\
CWE-415 & Double Free & 2 & 10 & 20.00 && 6 & 10 & 60.00 & $\uparrow$40.00\% \\
CWE-401 & Memory Leak & 2 & 8 & 25.00 && 3 & 8 & 37.50 & $\uparrow$12.50\% \\
CWE-22 & Path Traversal & 0 & 6 & 0.00 && 1 & 6 & 16.67 & $\uparrow$16.67\% \\
CWE-400 & Resource Exhaustion & 1 & 3 & 33.33 && 2 & 3 & 66.67 & $\uparrow$33.33\% \\
CWE-191 & Integer Underflow & 0 & 3 & 0.00 && 1 & 3 & 33.33 & $\uparrow$33.33\% \\
CWE-704 & Type Conversion Error & 0 & 2 & 0.00 && 1 & 2 & 50.00 & $\uparrow$50.00\% \\
CWE-264 & Access Control & 0 & 1 & 0.00 && 1 & 1 & 100.00 & $\uparrow$100.00\% \\
CWE-665 & Improper Initialization & 0 & 1 & 0.00 && 1 & 1 & 100.00 & $\uparrow$100.00\% \\
\bottomrule
\end{tabular}
\end{table*}

\noindent\textbf{\textit{Analyses.}} To systematically evaluate the CWE-specific effectiveness, Table~\ref{tab:cwe_results} compares the repair success rates between the base model (StarCoder + bug-fix data + vul-fix data) and our \vulkey approach across 15 CWE categories. These categories represent the CWE types for which the success rate changed among 69 CWE types in the test dataset.

\textbf{Our approach achieves repair rate improvements across all 15 CWE categories, demonstrating the universal effectiveness. } 
Experimental results highlight the importance of expert knowledge integration in \vulkey. First, the knowledge-driven generation mechanism significantly improves repair rates for common vulnerabilities, such as buffer errors (CWE-787/125), where explicit boundary check guidance leads to 11.11\% and 6.25\% EM gains, outperforming conventional CWE-label methods. Second, our three-level knowledge encoding enables precise fixes for memory safety issues, achieving 40.00\% EM improvement for double-free (CWE-415) and 10.34\% for use-after-free (CWE-416) by generating precise repair guidance of memory safety violations. Third, for previously unfixable cases (CWE-22/191/704/264/665), \vulkey attains 16.67–100.00\% success rates, indicating its ability to address subtle vulnerabilities missed by prior models. 
These findings confirm that our knowledge abstraction successfully translates CWE types into actionable repair patterns, boosting the AVR's effectiveness by providing explicit repair direction.

\noindent\textbf{\textit{Manual Analysis for Semantic Equivalence.}}
We manually assess patch semantic equivalence for the strongest baseline, StarCoder (tuned on bug and vulnerability data), and \vulkey. For each vulnerability instance, we select the top-5 candidate patches by Ratcliff/Obershelp similarity to the ground truth~\cite{ratcliff1988pattern}, and three independent authors ($>$5 years security auditing experience) verify semantic equivalence. We report inter-annotator agreement using Cohen's $\kappa$~\cite{cohen1960coefficient} (interpreted by Landis and Koch~\cite{landis1977measurement}); disagreements are adjudicated by the first and last authors. As shown in Table~\ref{tab:manual_analysis_results}, \vulkey attains a higher fraction of semantically equivalent patches than StarCoder, corroborating the gains observed under EM.

\begin{table}[t]
    \footnotesize
    \centering
    \caption{Results of Manual Semantic Equivalence Verification}
    \label{tab:manual_analysis_results}
    \renewcommand{\arraystretch}{1.10}
    \setlength{\tabcolsep}{6pt}
    \begin{tabular*}{\linewidth}{@{\extracolsep{\fill}}lcccc}
        \toprule
        \textbf{Model/Baseline} & \textbf{Total Instances} & \textbf{Semantically Eq.} & \textbf{Eq. (\%)} & \textbf{Cohen's $\kappa$} \\
        \midrule
        StarCoder + bug + vul & 435 & 159 & 36.5 & 0.90 (almost perfect) \\
        \vulkey & 435 & 188 & 43.2 & 0.84 (almost perfect) \\
        \bottomrule
    \end{tabular*}
\end{table}

\subsubsection{RQ2: What is the quality of our extracted/matched expert knowledge patterns?}
\ 
\newline
\textbf{\textit{Sampling-based Manual Validation.}}
To quantify the quality of our (Action, Key Element) patterns, we conduct a manual validation on the PrimeVul test set. We sample $n_e=100$ extracted and $n_m=100$ matched patterns via CWE-stratified random sampling (per-CWE cap of 5, fixed seed). For matched patterns we report both Top-1 and Top-10. Three independent authors ($>$5 years security auditing experience) label each pattern on: (Q1) Action-correctness (binary)---whether the Action matches the core edit operator in the ground-truth patch; (Q2) Key-Element correctness (binary)---whether the Key Element captures the core security mechanism of the patch; and (Q3) Actionability (three-level: noisy, partially useful, clearly actionable). We report accuracy for Q1/Q2 and the Q3 distribution from adjudicated labels. Inter-annotator agreement is the average pairwise Cohen's $\kappa$ from initial independent labels, interpreted per Landis and Koch~\cite{landis1977measurement}.

\begin{table}[t]
\centering
\caption{Results of Manual Knowledge Quality Validation.}
\label{tab:knowledge_quality_validation}
\footnotesize
\renewcommand{\arraystretch}{1.05}
\setlength{\tabcolsep}{4pt}
\begin{tabular*}{\linewidth}{@{\extracolsep{\fill}}lcccccccc}
\toprule
\textbf{Set} & $\mathbf{n}$ & \textbf{Action-corr. (\%)} & $\boldsymbol{\kappa_{\text{Q1}}}$ & \textbf{KE-corr. (\%)} & $\boldsymbol{\kappa_{\text{Q2}}}$ & \textbf{Actionability (L0/L1/L2)} & $\boldsymbol{\kappa_{\text{Q3}}}$ \\
\midrule
Extracted patterns & 100 & 97.0 & 0.71 & 93.0 & 0.68 & 2/4/94\textsuperscript{a} & 0.52 \\
Matched (Top-1) & 100 & 43.0 & 0.81 & 19.0 & 0.90 & 35/37/28\textsuperscript{a} & 0.53 \\
Matched (Top-10) & 100 & 80.0 & 0.75 & 58.0 & 0.72 & 51/93\textsuperscript{b} & 0.58 \\
\bottomrule
\end{tabular*}

{\footnotesize\raggedright
\textsuperscript{a}Level 0/1/2 distribution (\%); \textsuperscript{b}At least one Level 2 / Level 1+2 hit rate (\%).\\
$\kappa_{\text{Q1}}$, $\kappa_{\text{Q2}}$, $\kappa_{\text{Q3}}$: average pairwise Cohen's $\kappa$. Following Landis and Koch~\cite{landis1977measurement}: $0.41$--$0.60$ = moderate, $0.61$--$0.80$ = substantial, $0.81$--$1.00$ = almost perfect.\par}
\end{table}

Table~\ref{tab:knowledge_quality_validation} shows that our extracted patterns are generally high-quality and actionable: Action-correctness reaches 97.0\%, Key-Element correctness reaches 93.0\%, and 94\% of patterns are rated clearly actionable (L2). At inference time, the top-$k$ candidate mechanism substantially improves matching quality: Top-10 Action-correctness reaches 80.0\%, KE-correctness reaches 58.0\%, and the L1+L2 hit rate reaches 93\%, demonstrating that the multi-candidate strategy effectively supplies actionable repair guidance to the generator. Inter-annotator agreement across all dimensions reaches moderate to almost perfect levels.

\noindent\textbf{\textit{Top-$k$ Matching Quality (Precision@10/Recall@10).}}
We further report first-stage matching quality on the PrimeVul test set ($N=435$, $k=10$). A candidate is correct if it matches the ground-truth pattern (Section~\ref{sec:extraction}). Recall@10 ($=$ Hit@10) is the fraction of instances with at least one correct candidate in the top-10; Precision@10 is the per-instance fraction of correct candidates, averaged over instances. Overall (Action+Key Element) achieves 56.8\% Recall@10 / 16.4\% Precision@10; Action-only 82.3\% Recall@10 / 42.0\% Precision@10; Key-Element-only 64.6\% Recall@10 / 34.7\% Precision@10.

\begin{table*}[t]
\centering
\begin{minipage}{0.48\textwidth}
    \caption{Ablation Study of \vulkey's Core Components}
    \label{tab:ablation}
    \scriptsize
    \renewcommand{\arraystretch}{1.08}
    \setlength{\tabcolsep}{3pt}
    \begin{tabular}{@{}>{\raggedright\arraybackslash}p{0.45\linewidth}>{\raggedright\arraybackslash}p{0.24\linewidth}cc@{}}
    \toprule
    \textbf{Sub Model} & \textbf{Setting} & \textbf{Sample} & \textbf{EM} \\
    \midrule
    Full & \vulkey & 100 & 31.5 \\
    \midrule
    Expert Knowledge Extraction & \textit{w/o CWE type} & 100 & 26.4 \\
    \midrule
    \multirow{2}{*}{Code generation} & \textit{w/o Key element} & 100 & 27.6 \\
    \cmidrule(lr){2-4}
    & \textit{w/o Action} & 100 & 27.1 \\
    \bottomrule
    \end{tabular}
\end{minipage}
\hfill
\begin{minipage}{0.48\textwidth}
    \centering
    \includegraphics[width=0.74\linewidth]{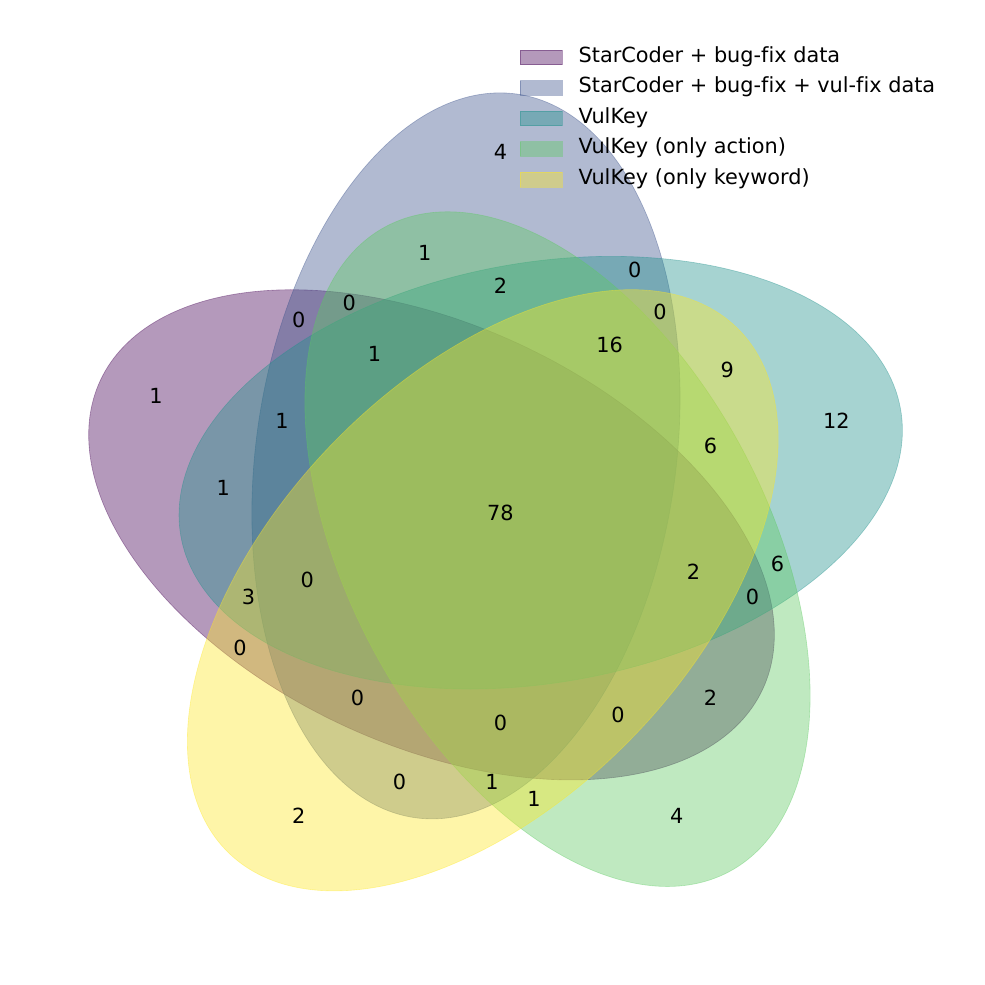}
    \captionof{figure}{Venn Diagram for Different Settings}
    \label{fig:venn}
\end{minipage}
\end{table*}

\subsubsection{RQ3: What's the contribution of each component in our method?}
\ 
\newline
\noindent\textbf{\textit{Setup.}} This RQ aims to investigate the contributions of the key designs of our approach. Specifically, we conduct ablation studies by removing one component at a time to assess the individual contributions of these key designs. In both the training and inference phases of Expert Knowledge Extraction, we ablate the CWE type and use only the vulnerable function as input to predict the action and key element. In the inference phase of Code Generation, we ablate either the key element or the action. It should be noted that this setting is fair, as the Code Generation model only includes the buggy/vulnerable function as input during training.

\noindent\textbf{\textit{Results.}} 
As shown in Table~\ref{tab:ablation}, CWE-type significantly boosts performance. The key element and action contribute equally. The full VulKey model's 31.5\% EM score demonstrates the necessity of integrating all components for optimal performance. The full VulKey model's unique sample coverage in the Venn diagram Figure~\ref{fig:venn} confirms their combined effectiveness.
Additionally, analyzing the Venn diagram reveals that using either the key element or action individually also results in some unique successful outcomes. This indicates that each component has its own strengths and can independently contribute to the repair process. 
Our efficient integration benefits from the strengths of both dimensions, achieving the optimal solution. Key elements provide fine-grained semantic guidance, while actions offer coarse-grained syntactic guidance. This combination allows the model to leverage detailed semantic cues alongside broader syntactic structures, facilitating a more comprehensive and effective approach to search for the best repair code.

\subsubsection{RQ4: What is the generalizability to another programming language?}
\ 
\newline
\noindent\textbf{\textit{Motivation.}} In the previous experiments, we mainly used C/C++ datasets for training and evaluation. To explore the generalizability to another language, we chose the Vul4J benchmark to evaluate the effectiveness of our work for the Java language.

\noindent\textbf{\textit{Setup.}} To facilitate a comprehensive comparison, we extracted the reported results from recent AVR studies~\cite{10.1145/3597926.3598135,ntr,10.1145/3597503.3639222}. Our analysis includes a comparison with the best results from the study~\cite{10.1145/3597926.3598135}, two state-of-the-art AVR tools, and one state-of-the-art template-based APR tool. In this section, we utilize the code generation model from previous experiments, while re-training the expert knowledge matching model of \vulkey using the Java dataset $X_1$. Depending on the evaluation result, we choose the two best-performing models StarCoderBase-15B and Qwen2.5-Coder-32B as the code generation model of \vulkey, denoted as $VulKey_{sc}$ and $VulKey_{qc}$.

\noindent\textbf{\textit{Results.}} 
As shown in Table~\ref{tab:full_results}, \vulkey achieves state-of-the-art performance with 18 successful repairs on Vul4J, representing 350\%, 100\%, and 64\% relative improvement over VulRepair (4 repairs), VulMaster (9 repairs), and NTR (11 repairs), respectively. The results demonstrate that our approach maintains strong cross-language generalizability when adapted with language-specific expert knowledge (trained on the Java dataset $X_1$). 

In addition, it also shows the consistent improvement in repair accuracy in different code generation models with our integration of expert knowledge. For example, $VulKey_{sc}$ improves StarCoder + Bug + Vul by 27.3\%/28.6\% relative improvement (11/14→14/18) and $VulKey_{qc}$ improves QwenCoder + Bug + Vul by 50\%/23.1\% relative improvement (6/13→9/16). The results demonstrate the cross-model generalizability of our approach.

\begin{table*}[t]
\centering
\caption{Performance of Different Approaches on Vul4J Dataset. Bug = Bug-fix data only, Bug+Vul = Bug-fix + Vul-fix data.}
\label{tab:full_results}
\setlength{\tabcolsep}{3pt} 
\tiny 
\begin{adjustbox}{max width=\textwidth}
\begin{tabular}{@{}lcccccccccccccc@{}}
\toprule
\multirow{2}{*}{Model} & \multicolumn{2}{c}{StarCoder} & \multicolumn{2}{c}{QwenCoder} & \multicolumn{2}{c}{CodeLlama} & \multicolumn{2}{c}{DSCoder} & \multicolumn{4}{c}{Baseline Models} & \multirow{2}{*}{$VulKey_{sc}$} & \multirow{2}{*}{$VulKey_{qc}$} \\
\cmidrule(lr){2-3} \cmidrule(lr){4-5} \cmidrule(lr){6-7} \cmidrule(lr){8-9} \cmidrule(lr){10-13}
Data Config & Bug & Bug+Vul & Bug & Bug+Vul & Bug & Bug+Vul & Bug & Bug+Vul & NTR & VulMaster & Codex-12B & VulRepair  \\
\midrule
Sample Size & 100 / 1000 & 100 / 1000 & 10 / 100 & 10 / 100 & 10 / 100 & 10 / 100 & 100 / 1000 & 100 / 1000 & 100 & -- & 10 & 100 & 100 / 1000 & 10 / 100 \\
Vul4J(35) & 8 / 13 & 11 / 14 & 6 / 11 & 6 / 13 & 4 / 11 & 8 / 12 & 7 / 8 & 7 / 11 & 11 & 9 & 6 & 4 & \textbf{14 / 18} & \textbf{9 / 16} \\
\bottomrule
\end{tabular}
\end{adjustbox}
\end{table*}

\section{Discussion}
\subsection{Failure Cases Root Cause Analysis}
\vulkey achieves a relatively low fix rate, largely due to the inherent complexity of certain vulnerabilities. We manually analyzed 100 failed cases by comparing the most similar predictions (via similarity-based matching) with the ground-truth fixes. Major failure sources include missing cross-function context (35 cases; \eg struct/union members defined outside the function scope), user-defined APIs (24 cases; difficulty recognizing and correctly invoking project-specific functions that deviate from standard library conventions), exact string/constant matching (10 cases; sensitivity to precise literals and constants), and complex logical conditions (9 cases; challenges with nested/multi-operator reasoning), with the remainder due to code-complexity mismatch (6; long or irregular edits), semantically equivalent but syntactically different patches (5; correct intent but mismatched surface form under EM), and other issues (11). Overall, these results suggest that \vulkey's primary limitations stem from insufficient global context awareness, inadequate adaptation to project-specific code patterns, and difficulties in handling complex logical constructs. However, these limitations do not diminish our core contribution, as our patterns can be seamlessly integrated into architectures incorporating global context collection to provide crucial security-related insights.

Based on this analysis, we outline several future directions. First, enrich global context by integrating repository-level context collection, such as cross-file dependency analysis and inter-procedural slicing, and feed retrieved definitions and usages into the repair model. Second, better adapt to project-specific code patterns by retrieving and injecting project-local API usages and exemplars, and exploring lightweight continual adaptation. Third, improve handling of complex logic by combining pattern-guided generation with verification- or constraint-guided reasoning, such as symbolic checks for path conditions, to reduce logical errors.

\subsection{Threats to Validity}
\textbf{Internal.}
One common issue with LLMs is data leakage. We mitigate this threat by using the open-source model StarCoder, which provides an official mechanism~\cite{starcoderopendata} to check for input code overlap with its training data. Using the official tool, we found only 13 of 435 PrimeVul test samples overlapping with StarCoder's training corpus. Importantly, \textsc{VulKey} correctly repaired only 2 of these overlapped cases, indices 55 and 338, suggesting that the impact of memorization is negligible. Furthermore, we evaluate our framework's effectiveness based on its incremental improvement over a base fine-tuned model, which inherently mitigates data leakage concerns. 

\textbf{External.}
Our external validity is threatened in three ways. First, for rare CWE types that are not covered in the training set, effective patterns cannot be matched through the CWE label alone. We partially mitigate this threat because the matcher can still leverage the vulnerable code itself to infer plausible repair patterns.
Second, our evaluation covers only C/C++ and Java, so the findings may not fully transfer to other programming languages. This risk is partly alleviated by the multilingual capabilities of modern code LLMs~\cite{codellama2024,starcoder2023,qwencoder2024,dscoder2024}, but broader cross-language validation is still needed.
Third, following prior works~\cite{fu2022vulrepair,vrepair,10.1145/3597503.3639222,10.5555/3766078.3766309,ntr}, we assume oracle vulnerability metadata (\ie identified buggy statements and known CWE type). In practice, these inputs come from imperfect upstream SAST tools (\eg CodeQL~\cite{codeql}, SonarQube~\cite{sonarqube}), which may reduce end-to-end effectiveness. We therefore view our results as an upper-bound evaluation and leave robustness to noisy localization/CWE signals to future work.

\section{Related Work}

\subsection{Program Analysis-Based Automated Vulnerability Repair}

Program analysis-based approaches are designed to handle only a few specific types of CWE. For example, LeakFix~\cite{10.5555/2818754.2818812}, MemFix~\cite{10.1145/3236024.3236079}, FootPatch~\cite{10.1145/3180155.3180250}, and SAVER~\cite{10.1145/3377811.3380323} are designed to address errors related to memory allocation and deallocation. This kind of work obtains an abstract representation of a program through modeling and performs verification and analysis on this representation. LeakFix performs verification and analysis on the control flow graph, and SAVER does verification and analysis across the object flow graph. They detect and fix memory errors by checking whether the safety conditions on the graph are violated. This straightforward modeling and analysis approach requires extensive expert knowledge to formulate detection and repair rules, making the development of such tools very time-consuming and labor-intensive. Currently, program analysis-based tools can only handle a very limited number of error types, and their effectiveness is suboptimal when faced with the diverse range of vulnerabilities present in real-world scenarios.

\subsection{Learning-Based Automated Vulnerability Repair}

Learning-based approaches initially treat the vulnerability repair task as a neural network translation task. For instance, SeqTrans~\cite{chi2020seqtrans} employs a Transformer-based machine translation model. Subsequently, VRepair~\cite{vrepair} employs a vanilla Transformer-based model with transfer learning to benefit from the large bug-fixing dataset. Later, with the advent of the era of LLMs, LCMs also entered the field of AVR. VulRepair~\cite{fu2022vulrepair} directly utilized CodeT5~\cite{wang2021codet5}, an LCM with the encoder-decoder architecture. Following this, VulMaster~\cite{10.1145/3597503.3639222} also adopted CodeT5. VulMaster achieves improved repair results by expanding the sources of input using several CWE type-specific typical vulnerability examples from the CWE website.
More recently, VulMatch~\cite{cao2024enhancing} retrieves repair patterns via nearest-neighbor lookup over clustered repair representations, and APPATCH, also known as A3P~\cite{10.5555/3766078.3766309}, follows an inference-time exemplar-based prompting paradigm by pairing model-generated vulnerability reasoning with original patches as exemplars.

In summary, while prior learning-based approaches have advanced the field, they share a common limitation in their handling of domain knowledge. Approaches like VulRepair treat knowledge as a simple textual prefix, failing to capture the underlying type-specific repair strategy. Others, like VulMaster, rely on rigid, concrete examples that struggle with generalization. Both VulMatch and APPATCH rely on patch-level representations or exemplars that inevitably retain project-specific identifiers and incidental logic, making retrieved guidance noisy. VulMatch further clusters and averages these representations for nearest-neighbor retrieval, which can blur security intent, and its capacity is bounded by the coverage of stored clusters. APPATCH conditions exemplar selection on free-form LLM-generated root-cause and strategy narratives together with an LLM match decision. This enlarges the matching space, challenges matching accuracy, and allows errors to cascade from reasoning to retrieval and synthesis. In contrast, we distill repair knowledge into discrete, multi-level (Action, Key Element) patterns organized by CWE; the symbols are recombinable across cases, and a dedicated matcher selects the most suitable pattern from this compact, structured space, yielding interpretable, stable, and generalizable guidance for the LLM.

\section{Conclusion}
In this work, we address the challenge of effectively integrating security-specific domain knowledge into LLM-based vulnerability repair. We introduced a three-level hierarchical abstraction for repair patterns, which captures the essential strategies for fixing vulnerabilities by combining CWE types with syntactic actions and semantic key elements. These syntactic actions point out specific edit operations like Insert Variable, and code elements represent the underlying security mechanism, with CWE type serving as a group indicator, thereby providing clear instructions for patch generation models.
This pattern abstraction is the core of our framework, \vulkey, which is built in two stages: expert knowledge matching and repair code generation. \vulkey first identifies the most relevant repair pattern for a given vulnerability and then uses this pattern to guide large code models in generating the patch. Our experimental results confirm the effectiveness of this method, showing that \vulkey achieves state-of-the-art performance and substantially outperforms existing baselines on all benchmarks.


\begin{acks}
This work was supported by the Research Grants Council of the Hong Kong Special Administrative Region, China (No. SRFS2425-4S03 of the Senior Research Fellow Scheme and No. CUHK 14209124 of the General Research Fund).
\end{acks}

\bibliographystyle{ACM-Reference-Format}
\bibliography{main}

\end{document}